\documentclass[aps,prb,twocolumn,groupedaddress,showpacs,10pt,amssymb,amsfonts]{revtex4-1}
\usepackage{amsmath,latexsym,hyperref,epsfig,subfigure,bm}

\def\beq{\begin{equation}}
\def\eeq{\end{equation}}
\def\be{\begin{equation}}
\def\ee{\end{equation}}

\renewcommand{\bf}{\mathbf}
\def\d{\partial}
\def\Tr{\mathrm{Tr}}
\def\tr{\hbox{tr}\,}

\def\cH{{\cal H}}
\def\rf#1{(\ref{#1})}
\def\rfs#1{Eq.~\rf{#1}}

\def\X{\bf{X}}
\def\dd{\bf{d}}
\def\chib{\bm{\chi}}

\def\aem#1{\marginpar{\small AE: #1}}

\begin{document}

\title{Bulk-boundary correspondence of topological insulators from their Green's functions}
\author{Andrew M.~Essin}
\author{Victor~Gurarie}
\affiliation{Department of Physics, CB390, University of Colorado,
Boulder CO 80309, USA}
\date{\today}

\begin{abstract}
Topological insulators are noninteracting, gapped fermionic systems which 
have gapless boundary excitations. They are characterized by topological 
invariants, which can be written in many different ways, including in terms 
of Green's functions.  Here we show that the existence of the edge states 
directly follows 
from the existence of the topological invariant written in terms of the 
Green's functions, for all ten classes of topological insulators in all 
spatial dimensions. We also show that the resulting edge states are characterized by their 
own topological invariant, whose value is equal to the topological invariant of the bulk insulator. 
This can be used to test whether a given model Hamiltonian can describe an edge of a topological insulator. 
Finally, we observe that the results discussed here apply equally well to interacting topological insulators, with
certain modifications. 
\end{abstract}

\maketitle

\section{Introduction}

The time-reversal invariant topological insulators of recent interest,
and the integer quantum Hall systems of longstanding 
interest (for recent reviews at a variety of levels, please see 
Ref.~\onlinecite{*HasanKane2010, *Moore2010, *QiZhang2010, *HasanMoore2011, *QiZhang2010-2}),
are now known to be just two elements in a classification table of all 
noninteracting fermionic systems,~\cite{Kitaev2009,Ryu2010}
which identifies ten symmetry classes of topological insulators. All of 
these systems are gapped in the bulk and possess a bulk topological invariant. They  are also supposed to have topologically 
protected gapless excitations at the boundary.  These gapless excitations are 
often taken as the most significant, even the defining, property of 
topological insulators.

A variety of arguments can be 
given supporting the existence of gapless edge states based on bulk 
properties, some more detailed (such as Laughlin's original argument 
designed for the integer quantum Hall effect~\cite{Laughlin1981}) and some 
more qualitative.  For example, consider the case of a model of 
noninteracting fermions that supports both a topologically trivial phase and 
a nontrivial phase, with the phase transition driven by varying a parameter 
in the system's Hamiltonian.  At the transition, corresponding to a special
value of the parameter, the excitation gap must close.
A boundary between two samples in the two phases can then be seen as a 
domain wall across which the parameter varies spatially through its special
value.  This ``spatial phase transition'' should then also result in gapless
excitations, which form in the vicinity of the special value.
%
%
Such arguments, while correct, tell us nothing about what kind of gapless edge excitations might form at the boundary of the topological insulator. 

Here we rely on the method of Green's functions to give a 
general, quantitative argument that
proves the existence of the edge states for all topological insulators. 
Moreover, our argument shows that the edge states are 
described by their own topological invariant,
analogous to the winding number of a vortex in a superfluid or to the 
charge of a particle as measured by Gauss' law 
(and similar to the defect 
invariants of Teo and Kane~\cite{Teo2010}), whose value must be equal to 
the invariant of the bulk insulator.  This edge invariant vanishes 
if the edge states are gapped, giving a simple way to see why they must be 
gapless.  The existence of this invariant gives us a tool to test whether a 
particular Hamiltonian can describe an edge theory, as it is 
straightforwardly calculable for any Hamiltonian under consideration. 

Before we derive this argument later in this article,
let us describe the edge topological 
invariant. A $d$-dimensional topological insulator that is translationally invariant,
so that the $d$-dimensional momentum ${\bf p}^d$ is a good quantum number,
possesses a $d$-dimensional (bulk) topological invariant 
$N_d$.~\cite{Ryu2010}   If the topological insulator has an edge, translation invariance 
in the direction perpendicular to the edge is lost, and the good quantum 
number is the $(d-1)$-dimensional momentum ${\bf p}^{d-1}$ parallel to the 
edge.
%
%
Let us take one of the 
components of this momentum, say 
$p_{d-1}$, and fix it at some large value $\Lambda$.  Although we expect the edge
to have
gapless excitations at some momentum (and so is not an insulator), 
the Hamiltonian at fixed $p_{d-1}=\Lambda$ is gapped
if $\Lambda$ is sufficiently large (as a function of the remaining $d-2$ 
momenta), 
and so describes 
a $(d-2)$-dimensional insulator.  We will show that this insulator is a topological insulator, with the topological invariant $N_{d-2}(\Lambda)$. Finally, we will show that 
\be \label{eq:theorem} 
N_d=N_{d-2}(\Lambda)-N_{d-2}(-\Lambda), 
\ee 
which constitutes the main result of this article.  

\begin{figure}[ht]
\includegraphics[width=2in]{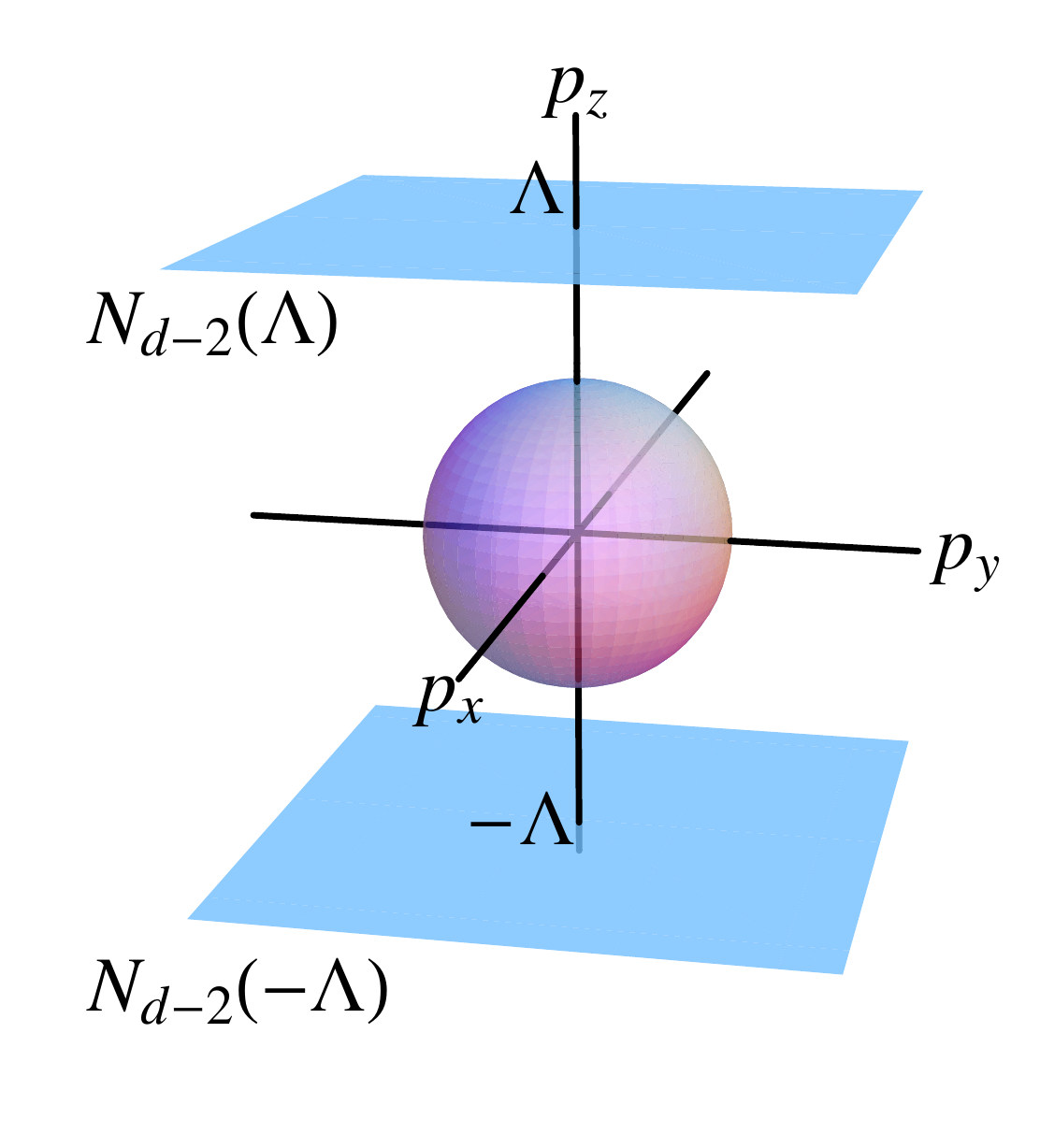} 
\caption{
The Fermi surface of radius $\mu/v$ of the topological insulator discussed in the text,
and the surfaces $p_z = \pm \Lambda$ on which the
edge topological invariant is computed. 
}
\label{fig:contour}
\end{figure}

As an example, consider the three-dimensional edge of a four-dimensional 
time-reversal invariant topological insulator with spin-orbit coupling. 
Suppose the excitations localized at the boundary are described by the 
Hamiltonian
\begin{equation} \label{eq:3dboundary} \cH = v \sum_{\alpha=x, y, z}  p_\alpha \sigma_\alpha -\mu, \end{equation}
where $\sigma_\alpha$ are the Pauli matrices and $v$ is Fermi velocity,  
which satisfies the appropriate symmetry (time-reversal invariance) $H(p) = \sigma_y H^*(-p) \sigma_y$.  
Its single-particle excitation spectrum is
\be \epsilon_{\pm}({\bf p}) = \pm v \sqrt{p_x^2+p_y^2+p_z^2}-\mu,
\ee and its zero energy excitations occur on a sphere of radius $\mu/v$, as shown in Fig.~\ref{fig:contour}. 
Let us see that this is indeed the edge of a topological insulator.  We fix 
$p_z=\pm \Lambda$ to obtain the two-dimensional Hamiltonian 
\begin{equation} \label{eq:2DHall} H_\pm = v \left( p_x \sigma_x + p_y \sigma_y \pm \Lambda \sigma_z \right) - \mu, \end{equation}
again as illustrated in Fig.~\ref{fig:contour}.
This Hamiltonian, understood as describing a two-dimensional system, is well known. For example, it has been
studied as a model of quantum Hall transitions.~\cite{Haldane1988,Ludwig1994}
The two-dimensional topological invariant of this Hamiltonian is known to 
be equal to its Hall conductance in units of $e^2/h$ (note that \rfs{eq:2DHall}, unlike \rfs{eq:3dboundary}, breaks time reversal invariance), and, assuming 
$\Lambda>\mu/v$, evaluates to  
\begin{equation} N_2(\Lambda)-N_2(-\Lambda)=1. \end{equation}
Therefore we conclude that this $H$ is a valid model of the boundary of a four-dimensional 
topological insulator.  This illustrates 
the usefulness of Eq.~\rf{eq:theorem}.

This example assumes an unphysically large bulk dimension, and 
the reader
may
doubt that it has any relevance for real-world physics.  However, it is 
known~\cite{Qi2008} that the experimentally realized time-reversal-invariant 
topological insulators in 2 and 3 dimensions can be viewed as descending
from a four-dimensional parent, simply by setting one or two of the momenta to zero.  Similarly, the three-dimensional boundary
just discussed can be seen as the parent of the boundary theories of those
real systems.  In particular, it follows from \rfs{eq:3dboundary} that the  boundary theory for the two dimensional time-reversal invariant topological insulator is
$H = v \, \sigma_x p_x - \mu$, while that of the three dimensional time-reversal invariant insulator is $H = v  \left( \sigma_x p_x + 
\sigma_y p_y \right)-\mu$. 
We will return to this ``dimensional reduction'' later in this article. 


The Green's function formalism we use is quite powerful: it remains 
applicable even when interactions are added to the system, since Green's 
functions exist for interacting as well as for noninteracting systems. Here one must comment that the power of topological invariants
often lies in the fact that they represent the response of noninteracting systems to external electromagnetic field. Once the interactions are turned on,
the topological invariant written in terms of interacting Green's functions  may or may not represent the response, and this could be considered an impediment to using the
Green's function formalism to describe interacting topological insulators. 
However, the main result, Eq.~\eqref{eq:theorem}, 
is correct 
regardless of the presence of interactions as will be clear from its derivation.  
If one takes as
the most physically relevant property of topological insulators is the existence of edge states, 
then
since
the relation Eq.~\eqref{eq:theorem} relates the topological invariant to the property of the edge, this relation, and not the response, can in principle be taken as a 
starting point of the application of this formalism to the interacting topological insulators.

Yet the interpretation of 
Eq.~\eqref{eq:theorem} may change once the interactions are turned on. 
As discussed in a recent paper by one of us,~\cite{Gurarie2011} in 
the presence of interactions the (eigenvalues of the) Green's functions may have zeroes localized 
at the edge, in addition to the poles that indicate the usual edge 
states.  From the point of view of the topological invariants, zeroes are 
similar to poles,
and can result in a nonzero edge invariant even in the absence of 
zero-energy edge states.\cite{*[{Recently, }] [{, have pointed
out that in a noninteracting system with an unbounded spectrum the same
thing can happen.  We assume that the Hamiltonian is finite and periodic,
as in the low-energy states of a crystal.}] SilaevVolovik2011} In fact, it can be shown that the recent result 
of Ref.~\onlinecite{Fidkowski2010}, where a model topological insulator 
loses its edge states in the presence of interactions, is due to the 
replacement of the edge states by the zeros of the Green's 
function.~\cite{GurarieEssin2011}
The study of the effects of interactions using the formalism developed here 
appears to be a useful direction of further research, but goes beyond the 
scope of this article.  In what follows we mostly assume that the fermions do not interact, although throughout this article we point out what exactly changes if
interactions are taken into account. 

To connect this article to previous publications, we remark that the relation \rfs{eq:theorem} was discussed for $d=2$ in Ref.~\onlinecite{VolovikBook1}. Subsequently,
it was extended for $d=1$ and discussed in the presence of interactions in Ref.~\onlinecite{Gurarie2011}. Here we extend it for all topological insulators, in any number
of dimensions. 

We proceed as follows: in section \ref{sec:nosym} we derive Eq.~\eqref{eq:theorem} for systems with
no symmetry other than translational (class A in the Altand-Zirnbauer 
classification \cite{Zirnbauer1996,Altland1997}), which only have integer-valued topological invariants in even spacial dimensions. 
Then in section \ref{sec:trph} we describe the effects of discrete symmetries on the invariants; and 
then describe how these results imply the relation between bulk 
and surface physics for $\mathbb{Z}_2$ topological systems. Finally in section \ref{sec:chiral} we discuss systems with chiral symmetry. 
Appendix \ref{sec:derivation} contains the derivation of a key step in 
our derivation of Eq.~\eqref{eq:theorem}, and Appendix
\ref{sec:chiralone} presents a more complete discussion of 1D topological insulators.
%


\section{Topological insulators without any symmetries}
\label{sec:nosym}
We start with a translationally invariant  topological insulator 
with a single particle 
Green's function $G_{\alpha \beta}(\omega, {\bf p})$.  
If interactions are absent, the insulator can be
described by the Hamiltonian
\begin{equation} 
H = \sum_\bf{p} \cH_{\alpha\beta}(\bf{p})  \, 
\hat{a}^\dag_{\alpha\bf{p}} \hat{a}_{\beta\bf{p}}.
\end{equation}
Here $\hat{a}^\dag_{\alpha\bf{p}}$ creates a fermion with
momentum $\bf{p}$ in $d$ dimensions, and the species label $\alpha$ runs
over the spin, bands, as well as particle/hole space for a superconductor (summation on indices should be always understood).  In that case, the Green's function is simply given by
\begin{equation} \label{eq:greendef} 
G(\omega,\bf{p}) = [i\omega - \cH(\bf{p})]^{-1}.
\end{equation} 
Once interactions are switched on, simple expressions of this kind are no longer available, but Green's function can still be defined in a standard way (see, for example, \onlinecite{FetterWaleckaQTMPS}).

The bulk topological invariant is
\beq \label{eq:invone} 
N_{d} = C_{d}\, \epsilon_{a_0 \dots a_d} \!\int\! 
d\omega d^dp \, \tr   G^{-1} \d_{a_0} G \dots  G^{-1} \d_{a_d}  G,
\eeq
where $a$ runs over $(\omega,\bf{p})$ , 
and $\epsilon$ is fully antisymmetric tensor, with
$\epsilon_{\omega p_1 \dots p_d}=+1$
(the other nonvanishing components are obtained by permutation).  The 
constant $C_{d}$ is given by~\cite{Ryu2010}
\beq
C_{d} = -(2\pi i)^{-(d/2)-1}(d/2)!/(d+1)!
\eeq
The spatial dimension $d$ is even, as $N_d$ vanishes when $d$ is odd 
(by antisymmetry 
of $\epsilon$ and cyclicity of the trace $\tr\!$); we will later introduce
a different expression, valid for chiral systems in odd dimensions, that we
denote by the same symbol.

This topological invariant always evaluates to an integer, and gives the Hall conductance in two-dimensional space, \cite{Niu1985} at least in the absence of interactions --- it remains a topological quantity (an integer) even with interactions.
This is true simply 
because it measures
the winding of the map from a ($d+1$)-dimensional space $(\omega, {\bf p}^d)$ to a space of matrices $G$; it is known that such a map can be topologically nontrivial if $d$ is even.
Moreover, it is clear that in the absence of interactions, to change the value topological invariant one needs to deform the Green's function in such a way that either $G$ or $G^{-1}$ becomes singular. For noninteracting systems, this is only possible if $\omega=0$ and $\cH$ has zero energy states as
follows from Eq.~\rf{eq:greendef};  this results in infinite $G$ at $\omega=0$ and the closing of the gap in the spectrum. For interacting systems $G$ itself can acquire zero eigenvalues at $\omega=0$,~\cite{Gurarie2011} so that $G^{-1}$ is singular; thus interacting systems can change the value of their invariant Eq.~\rf{eq:invone} without ever closing the gap. 
All of this matches known results. 

Now we introduce some technology, inspired by Volovik.\cite{VolovikBook1}  
Consider a $d$-dimensional system with 
a domain wall of dimension $d-1$; the Hamiltonian varies 
in the direction perpendicular to the domain wall in such a way that far from the domain wall the Hamiltonian describes a topological insulator
with the invariant $N_d^R$ or $N_d^L$, on the right or left side of the domain wall (what this means precisely will be defined below). 
Since there are only $d-1$ good momenta, it is described by the mixed 
Green's function $\tilde G(\omega,\bf{p}^{d-1};s,s')$, where $s$ is the 
coordinate normal to 
the boundary and is effectively another matrix index.
A Fourier transform 
with respect to $s-s'$ produces the Wigner (-Weyl) transform 
${G}(\omega,\bf{p}^d,\bar{s})$,~\cite{Kamenev2004,Littlejohn1986}
with $\bar{s}=(s+s')/2$ and $p_d$ the  momentum conjugate to $s-s'$.
Far from the domain wall, this will become $\bar s$-independent and 
coincide with the bulk Green's function (that is, with translation invariance
the Wigner and Fourier transforms are the same).

Importantly, two objects exist which can be interpreted as the inverse of 
$G$.  The ``true'' inverse 
satisfies
\begin{equation}  \label{eq:tildeK}
\sum_\beta \int ds' \tilde K_{\alpha \beta} ( s, s') \, \tilde G_{\beta \gamma} ( s',s'')  = \delta_{\alpha \gamma}
\delta(s-s'')
\end{equation}
(in this expression the $\omega$ and ${\bf p}^{d-1}$ dependence of $\tilde K$ and $\tilde G$ was suppressed for brevity). 
For noninteracting systems 
\begin{equation} 
\tilde K (\omega, {\bf p}^{d-1}; s, s')
= i\omega-\cH (\omega, {\bf p}^{d-1}; s, s'),
\end{equation}
but for interacting systems one should rely solely on Eq.~\rf{eq:tildeK} to calculate $\tilde K$. 
At the same time,
matrix (or local) inverse ${G}^{-1}$ satisfies
\begin{equation}
{G}^{-1}_{\alpha \beta} (\omega,\bf{p}^d,\bar{s}) \, 
{G}_{\beta \gamma} (\omega,\bf{p}^d,\bar{s}) = \delta_{\alpha \gamma}.
\end{equation} 
With translation invariance along $s$, 
${K} = {G}^{-1}$, but in the presence of a domain wall $K \neq G^{-1}$ [here $K(\omega, {\bf p}^d, \bar s)$ is the Wigner transform of $\tilde K(\omega, {\bf p}^{d-1}; s, s')$]. 

With these tools, we define the ($d+2$)-dimensional vector 
\beq \label{eq:vector} 
n_{a_0} = C_{d} \, \epsilon_{a_0 a_1 \dots a_{d+1}} 
\tr G^{-1} \d_{a_1} G \dots  G^{-1} \d_{a_d} G 
\eeq 
in the space $(\omega,\bf{p}^d,\bar{s})$.  
Remarkably, the divergence of this vector is zero except
where $G$ (or $G^{-1}$ if interactions are present) is singular, as can be checked by direct differentiation. 
This is completely analogous 
to the
electric field in electrostatics with point charges or charged surfaces.  
Here, the sources of $\bf{n}$ are singularities (or zeros) of ${G}$.  The 
analogy to electrostatics is productive; we will essentially measure the 
charge of a singularity with Gauss' law, integrating the flux of $\bf{n}$
over a surface that surrounds the charge.

For the rest of this section, for simplicity we will always refer to these as ``singularities" having in mind that in the absence of interactions $G$ cannot have zeros.
We will also keep in mind that if interactions are present, singularities will imply either infinite or zero $G$. 

The
singularities of $G$ may occur at multiple points  of the $d+2$ dimensional space 
or on surfaces in the space spanned by $(\omega, {\bf p}^d, \bar s)$. We denote these $f_i$. 


Because $G(\omega,\bf{p}^d,\bar{s})$ reduces to the bulk Green's function far from the boundary (domain wall), the definitions Eqs.~\eqref{eq:invone} and \eqref{eq:vector} mean that we can compute $N_d$ as the flux of $\bf{n}$ through the surface $\bar{s}=L$ for large $L$.  That is,
the difference between topological 
invariants on either side 
(right $N_d^R$ and left $N_d^L$, say) of the boundary, 
\beq \label{eq:topinvdif}
N_d^R - N_d^L = \int\! d\omega d^dp\, 
[n_{\bar{s}}(\bar s = L) - n_{\bar{s}}(\bar s = -L)],
\eeq
is just the flux of $\bf{n}$ through the combined surface 
$(\omega,\bf{p}^d,\pm L)$, shown in Fig.~2. 
\begin{figure}[ht]
\hspace{-20pt} \includegraphics[width=2in]{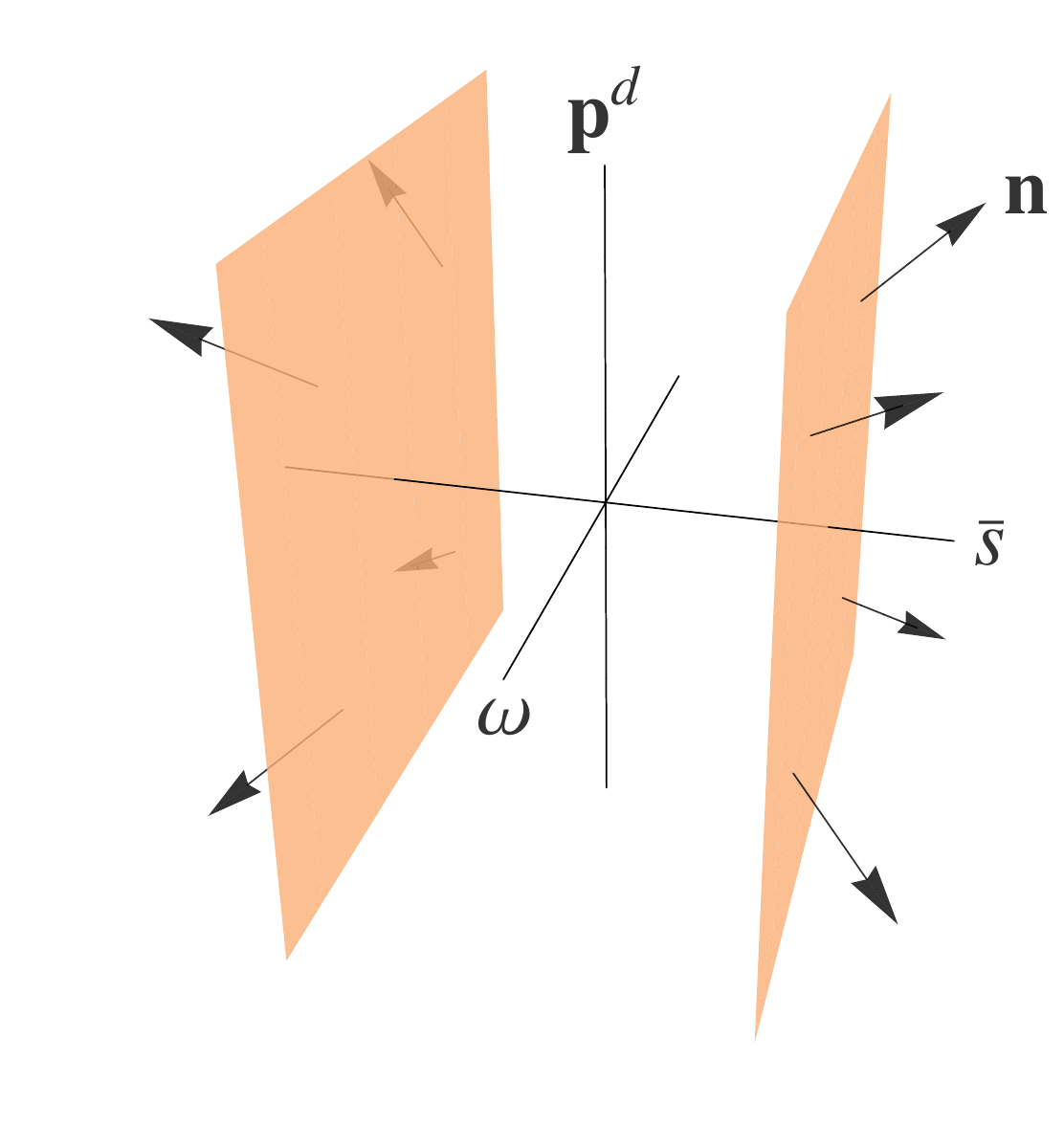} 
\caption{
The surfaces $\bar{s}=\pm L$ on which $N_d$ is computed as the flux
of $\bf{n}$.
}
\label{fig:contour1}
\end{figure}
On the other hand, 
and this is the crucial point, 
because $\d_a n_a = 0$ we can use 
\emph{any surface we like}
to compute $N_d^R - N_d^L$
so long as it 
encloses the singular surfaces of $G$. Denoting by $S_{f_i}$  any 
($d+1$)-dimensional 
``Gaussian" surface (for example, a sphere) surrounding $f_i$ we find
\beq \label{eq:nflux}
N_d^R - N_d^L = \sum_{i} \int\! \bf{dS}_{f_i} \cdot \bf{n} .
\eeq
Eq.~\eqref{eq:nflux} constitutes the first half of our argument.\cite{*[{Note that this result resembles the formalism of }] [{ for point defects.}] Santos2010}

To relate the flux of ${\bf n}$ to the edge properties of the system, we construct a $d$-dimensional vector $\bf{r}$ out of the mixed
Green's function $\tilde G(\omega,\bf{p}^{d-1};s,s')$, which allows us to 
define a topological invariant 
that directly captures the behavior of the edge states. 
%
%
We define
\be \label{eq:vectored}
r_{a_0} = C_{d-2} \, \epsilon_{a_0 \dots a_{d-1}}
\Tr\, \tilde K \circ \d_{a_0} \tilde G \circ\dots\circ \tilde K \circ \d_{a_{d-1}} \tilde G.
\ee
The convolution $(\tilde A\circ \tilde B)(s,s'') = \int\!ds' \tilde A(s,s') \tilde B(s',s'')$ that appears here is 
really
a generalization of matrix multiplication to functions of two coordinates $s,s'$.  That is, the coordinates $s,s'$ are
treated as matrix indices of the functions $\tilde A$ and 
$\tilde B$, and $\tilde G$; 
of course, these functions
also carry the ordinary matrix indices 
$\alpha, \beta$ that label fermion species, \textit{etc}, and which are being summed over in the usual way.
Then $\Tr \, \tilde A \equiv \int\! ds \, \tr\! \tilde A(s,s)$ is just
the trace for these generalized matrices.  The index $a$ here runs over 
$(\omega, {\bf p}^{d-1})$, the space of the vector $\bf r$.


The vector ${\bf r}$, just like the vector $\bf{n}$ above, is 
divergence-free, except at singularities of the mixed Green's function
$\tilde{G}$.
These singularities can also be interpreted as ``electrostatic'' sources for the  
vector ${\bf r}$ emanating from them. 

It is clear that the singular surfaces $f_i$
--- the sources of ${\bf n}$ 
--- 
become the sources of ${\bf r}$ as well when projected from the 
$(d+2)$-dimensional space $(\omega, {\bf p}^d, \bar s)$ onto
the $d$-dimensional 
edge
space $(\omega, {\bf p}^{d-1})$. 
Since the Green's function 
$\tilde G=\left[ i\omega - \cH \right]^{-1}$ can only be singular where 
$\omega=0$ and $\cH$ has zero energy eigenstates,  these sources 
are confined to $\omega=0$ 
and form surfaces
(or points) in the 
$(d-1)$-dimensional 
space spanned by ${\bf p}^{d-1}$,
which we 
identify with the edge Fermi surfaces 
and Dirac points 
and denote $F_i$ (having in mind that in the presence of interactions surfaces and points of zero $\tilde{G}$ play an equivalent role).


Remarkably, we can prove the following statement:
\be \label{eq:rnconnection}
\sum_{i} \int\! \bf{dS}_{f_i} \cdot \bf{n} 
= \sum_{i} \int\! \bf{dS}_{F_i} \cdot \bf{r}.
\ee
Here $S_{F_i}$ is the $(d-2)$-dimensional surface surrounding the Fermi 
surface (or Dirac point) $F_i$. 
The proof of this statement is given in Appendix \ref{sec:derivation} and is quite involved 
(although for $d=2$ the proof is significantly simpler; it is given in 
Ref.~\onlinecite{VolovikBook1}).  It relies on the 
approximation of the domain wall being smooth.  However, the corrections to 
this equation form a series expansion in powers of the gradient 
$\d_{\bar s} G$.  Since both sides of this equation
are integers due to Eq.~\rf{eq:nflux}, small corrections to it must vanish. 

It follows from Eqs.~\rf{eq:nflux} and \rf{eq:rnconnection} that
\beq \label{eq:bulkbound}  
N_d^R-N_d^L 
=  \sum_{i} \int\! \bf{dS}_{F_i} \cdot \bf{r}.
\eeq
In words, this says that the Fermi surfaces (and Dirac points) on the edge
are characterized by ``topological charges,'' or fluxes of ${\bf r}$ 
emanating from these surfaces and points.  This charge is equal to the 
difference in the bulk topological invariants on either side of the 
boundary.  Eq.~(\ref{eq:bulkbound}) is the quantitative 
statement of the fact that gapless excitations are required 
at a boundary
between bulk insulators with different values of the topological 
invariant $N_d$ (in the presence of interactions, there can be zeroes instead \cite{Gurarie2011}).  Indeed, in the absence of such excitations, 
${\bf r}$ is 
divergence-free everywhere and its flux is always zero. 

\begin{figure}[ht] 
\hspace{-20pt} 
\includegraphics[width=2.4in]{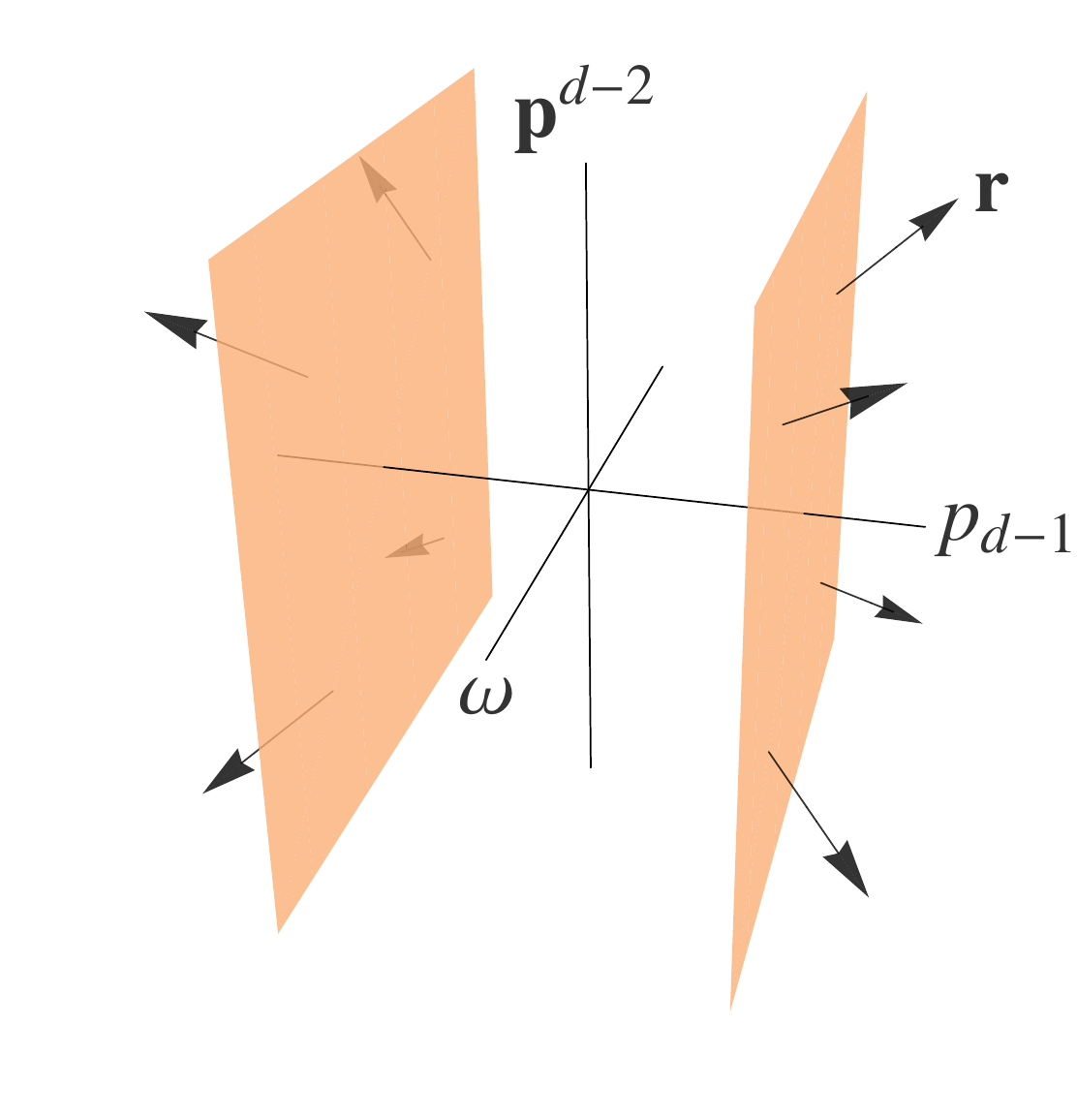} 
\caption{ \label{fig:rflux}
The surfaces $p^{d-1}=\pm \Lambda$ on which $N_{d-2}$ is computed as the flux
of $\bf{r}$.
}
\end{figure}

What remains is to interpret this charge as a difference of topological 
invariants, as discussed in the beginning of this article.  
So long as the 
Fermi surface does not traverse the Brillouin zone in the $p_{d-1}$ 
direction, 
we can choose $S_F$ to be the surfaces 
$(\omega, {\bf p}^{d-2}, \pm \Lambda)$ for suitable $\Lambda$, instead of 
the spheres closely surrounding the Fermi surface(s) that we assumed until 
this point [see Fig.~\ref{fig:rflux}].  

The flux through $S_F$ is then the difference of the fluxes of 
${\bf r}$ through these two surfaces.  In turn, those fluxes can be reinterpreted as the difference $N_{d-2}(\Lambda)-N_{d-2}(-\Lambda)$, where
\begin{equation} 
N_{d-2} (p_{d-1}) = \int\! d\omega\, d^{d-2} p \, r_{p_{d-1}} 
\end{equation}
is a $(d-2)$-dimensional topological invariant calculated in the space 
$(\omega, {\bf p}^{d-2})$ with $p_{d-1}$ fixed, as can be verified with
Eq.~\rf{eq:vectored}.  The main result of this article Eq.~\rf{eq:theorem} (where $N_d^L$ is assumed to be 0 for simplicity) immediately follows.

In the simplest case $d=2$ this result was derived in 
Ref.~\onlinecite{VolovikBook1}, where $N_0(\Lambda)-N_0(-\Lambda)$
was interpreted as a number of energy levels crossing zero as the momentum 
along the one-dimensional edge changes from $-\Lambda$ to $\Lambda$, corresponding to the standard picture of edge states
in the integer quantum Hall effect. In higher number of even dimensions $d$, this result's interpretation is more complicated, but 
the result is nonetheless useful,
as in the example given in the Introduction.  

This concludes the discussion of topological insulators without symmetries (systems of class A). 

\section{Topological insulators with time-reversal or particle-hole symmetry}
\label{sec:trph}
Now let us
apply this formalism to systems in the nine remaining classes of topological insulators with
symmetry, beginning from non-chiral insulators.  Nonchiral insulators, termed classes AI, AII, C and D, are those
which possess either time-reversal or particle-hole symmetry (it can be
helpful to refer to the ``periodic table'' of topological insulators in 
Refs.~\onlinecite{Kitaev2009,Ryu2010} for the discussion of this section
and the next).
We have seen that with no symmetry (class A), bulk insulators are
classified by an integer in even dimensions.  The same invariants describe the other nonchiral classes  in even dimensions. However the presence of discrete symmetries can force them to vanish in some dimensions, as follows.  
Classes AI and AII represent time-reversal ($T$) invariant systems with integer and 
half-integer spin respectively, and the Green's function satisfies 
the constraint
\begin{equation} \label{eq:tri}
G(\omega,\bf{p}, \bar s) 
= U_T^\dagger \,  G^T(\omega,-\bf{p}, \bar s) \, U_T, \end{equation}
where $U_T$ is a unitary matrix such that $U_T^* U_T = \epsilon_T=+1$ for 
AI and $\epsilon_T=-1$ for AII ($G^T$ is the transposed Green's function).
The same constraint holds for $\tilde{G}$.  Putting this into $N_d$, 
relabeling $\bf{p}\rightarrow-\bf{p}$, taking the transpose, using 
cyclicity, and relabeling indices, we see that this forces $N_d = 0$ when 
$d=4n+2$ (an anticyclic permutation is odd in $d=4n+2$).
Similarly, classes C and D have particle-hole ($C$) symmetry
\begin{equation} \label{eq:phi}
G(\omega,\bf{p}, \bar s) 
= - U_C^\dagger \,  G^T(-\omega,-\bf{p}, \bar s) \, U_C,
\end{equation}
where $U_C^* U_C=\epsilon_C=+1$ for class D and $\epsilon_C=-1$ for 
class C.  An additional minus sign in the integrals is generated by 
$\omega \rightarrow -\omega$, so that $N_d$ is nonzero only in spaces 
of dimension $d=4n+2$. All of this matches what was established elsewhere 
using a different language.~\cite{Ryu2010}

The conclusion here is that if we are in a spatial dimension where the topological invariant
can be nonzero, the bulk-boundary correspondence Eq.~\rf{eq:theorem} applies just as it does for systems without any symmetry. 
One might worry that the relation Eq.~\rf{eq:theorem} involves not only $N_d$ but also $N_{d-2}$; as we just saw, in systems with symmetries if $N_d$ is
nonzero, $N_{d-2}$ appears to be zero. However, in Eq.~\rf{eq:theorem} $N_{d-2}$ is calculated with $p^{d-1}$ fixed at some value $\Lambda$, so
the Green's function no longer satisfies Eq.~\rf{eq:tri} or \rf{eq:phi}. Thus both sides of Eq.~\rf{eq:theorem} can be nonzero, as they should. 

Let us now account for the appearance of $\mathbb{Z}_2$ 
invariants, which in time-reversal-invariant insulators which belong to class AII appear for example in $d=2$ and $d=3$.  To explain these, we appeal to the dimensional reduction 
picture 
of Qi, Hughes, and Zhang,~\cite{Qi2008} in which the physical 
system is considered as embedded within a space of higher dimension, 
with one or two of the momenta being fictitious additional parameters in 
the Green's function. The real physical system corresponds to the 
fictitious momenta set to zero. Under $T$ (or $C$) 
almost every point in the momentum space has an image at the
opposite momentum, the exceptions being the time-reversal-invariant (TRI)
points. 
As a result, if a Green's function has a point singularity at a non-TRI 
point, this contributes 2 to \rfs{eq:bulkbound}, 
since each point and its image contribute. So, if $N_d^R-N_d^L$ is 
odd and all singularities are points, 
some of the singular points must be located at the TRI points in the momentum 
space. Finally, it is possible to make sure, by choosing the appropriate 
extension of $G$ to the unphysical momenta, that all singular TRI points 
are in the physical space.  This leads to the conclusion that if the 
topological invariant in the extended space is
an odd integer, 
there must be singular points, or as we saw edge states, in the physical 
(reduced-dimension) theory. 

The same arguments apply if, instead of a Fermi 
point, there is a Fermi surface surrounding a TRI point.  The case of 
a surface is also crucial for showing that generally the dimensionality 
cannot be reduced by more than 2. 
Indeed, generalizing
a result due to Ho\v{r}ava,~\cite{Horava2005}
generic topological insulators characterized by integer 
invariants have two-dimensional Fermi surfaces at the boundary.  A Fermi 
surface (spherical, for simplicity) centered on zero is defined by the  
equation $p_x^2 + p_y^2 +p_z^2=p_F^2$, with the remaining momenta, if any, 
being arbitrary.  This can be dimensionally reduced by one or two by setting 
$p_z=0$ or $p_y=p_z=0$. However, to reduce the dimensionality by more than 2 
requires setting $p_x=p_y=p_z=0$, which eliminates gapless excitations. 

\begin{figure}[ht] 
\hspace{-20pt} 
\includegraphics[width=2.4in]{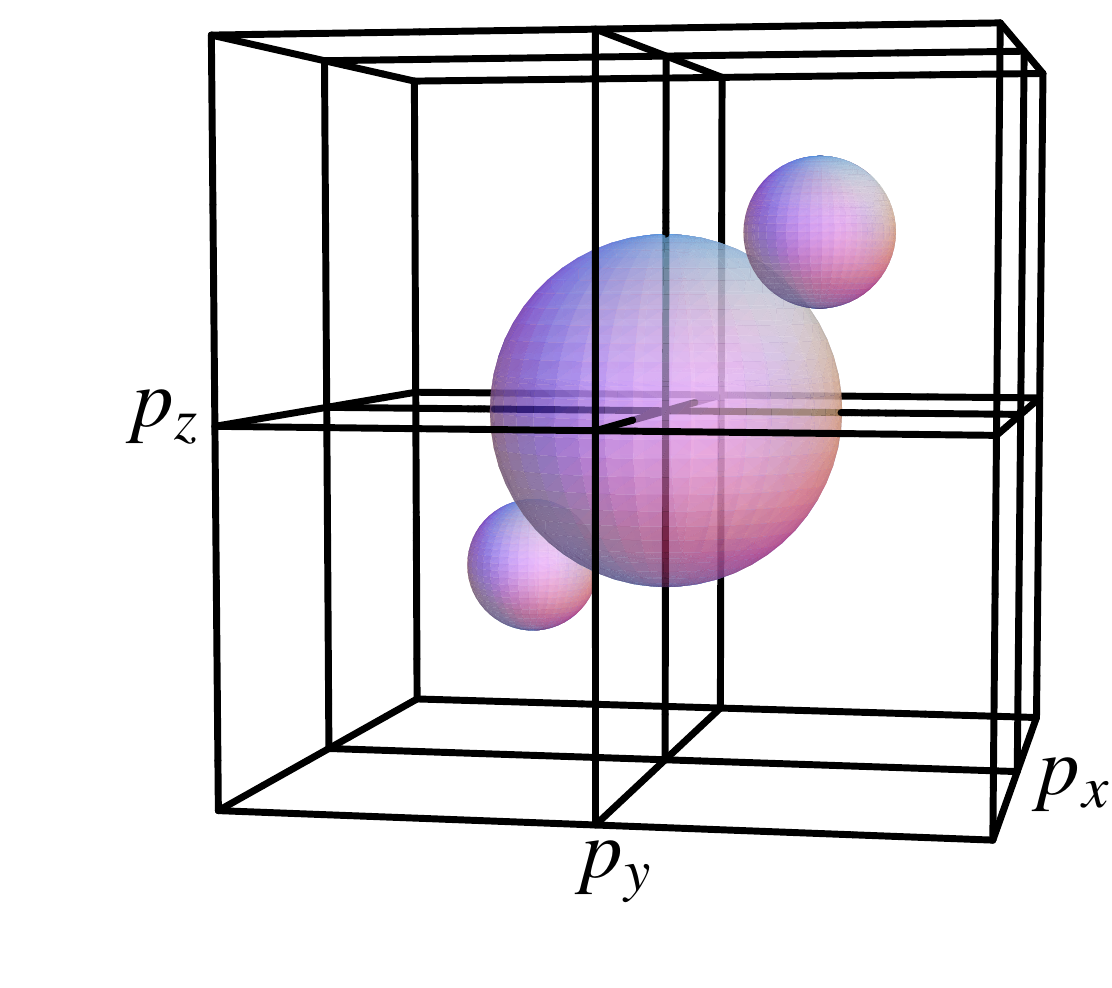} 
\caption{ \label{fig:threesurfaces}
Example of an edge Brillouin zone containing a time-reversal-invariant, multicomponent Fermi surface.  The two smaller 
spheres each contribute to the three-dimensional edge topological invariant,
but they do not intersect the time-reversal invariant planes of the Brillouin zone and so do 
not contribute to dimensionally reduced edges.
}
\end{figure}

An example of two-dimensional Fermi surfaces located at the 
three-dimensional edge of a four-dimensional topological insulator with 
time-reversal invariance is shown in Fig.~\ref{fig:threesurfaces}. 
Three Fermi surfaces are depicted, giving $N_4=N_{2}(\Lambda)-N_2(-\Lambda)=3$. At the same time, two of those Fermi surfaces located at opposite momenta play no role if dimensions are reduced by setting, for example, $p_z=0$.  The remaining Fermi surface centered on the origin still yields zero energy excitations even if $p_z=0$, thus in this example the edge state survives in lower dimensions. On the other hand, if $N_4$ were even, then the central Fermi surface would necessarily be absent (or there would be a pair that could be deformed off the $p_z=0$ plane), and there would be no edge excitations in lower dimensions. Here, as everywhere else, in the presence of interactions, the Fermi surfaces could be the surfaces of zero energy excitations as well as surfaces of zeroes of Green's functions. 

Note that if one breaks time reversal invariance by adding appropriate terms in the Hamiltonian, the excitations no longer have to be symmetric under reflection of momenta. Then the central Fermi surface can move off the center of the Brillouin zone and no longer contribute to the lower dimensional edge excitations. This is the mechanism by which breaking time reversal invariance removes edge states in lower dimensions (but not in the original 3-dimensional edge where edge excitations survive even if time reversal invariance is broken).

Finally, the dimensional reduction gives a trivial result in some 
dimensions because the bulk invariant $N_d$ takes only even values for the
dimensionally extended system, which washes out the $\mathbb{Z}_2$ 
structure.

These arguments conclude our discussions of topological insulators in classes AI, AII, C and D. 


\section{Topological insulators with chiral symmetry}
\label{sec:chiral}
We also want to show that analogous results hold for the classes
with chiral, or sublattice ($S$), symmetry (AIII, BDI, CII, CI and DIII).
In these systems there is a matrix $\Sigma$ such that 
\be G(\omega, {\bf p}, \bar s) = - \Sigma \,  G(-\omega, {\bf p}, \bar s) \, \Sigma, \quad \Sigma^2=1.\ee
The bulk invariant for chiral systems, analogous to \rfs{eq:invone},
can be written as
\beq \label{eq:topinvchialt} 
\hspace*{-7pt}
{N}_d = \frac{C_{d-1}}{2} \epsilon_{a_1 \dots a_{d}}  \!\!
\int\!\! d^d p \, \tr  \Sigma \, 
 G^{-1}\d_{a_1}  G \dots G^{-1}\d_{a_{d}}  G,
\eeq
where $G$ is evaluated at $\omega=0$, $a$ runs over the  components 
of $\bf{p}$, and the bulk dimension $d$ is now odd.~\cite{Volovik2010} 
Unlike the even-dimensional topological invariant \rfs{eq:invone}, this expression does not treat momenta and frequency in a symmetric fashion and in fact
is derived directly from the topological invariant written in terms of the Hamiltonian in Ref.~\onlinecite{Ryu2010} by replacing $\cH \rightarrow \left. G^{-1} \right|_{\omega=0}$. 
An expression for $N_d$ with $d$ odd which involves integration over frequency and momenta is also possible,~\cite{Gurarie2011} however for the purpose of this article it is not needed.

Expressions strictly analogous to Eqs.~(\ref{eq:vector}), 
(\ref{eq:nflux}),  and (\ref{eq:vectored}) can be 
formulated 
for the chiral case, leading to a relation
just like the bulk-boundary correspondence \rfs{eq:bulkbound}.
In particular,
\beq \label{eq:vector-chi} 
{n}_{a_0} = \frac {C_{d-1}} 2 \, \epsilon_{a_0 \dots a_{d}} 
\tr \Sigma \, G^{-1} \d_{a_1} G \dots G^{-1} \d_{a_d} G,
\eeq 
where $a$ runs over $({\bf p}^d, s)$, is a $(d+1)$-dimensional 
divergence-free vector, while
\beq \label{eq:vectored-chi}
{r}_{a_0} = \frac {C_{d-3}} 2 \, \epsilon_{a_0 \dots a_{d-2}}
\Tr\, \Sigma \, \tilde K  \circ \d_{a_1} \tilde G \circ \dots\circ \tilde K \d_{a_{d-2}} \tilde G
\eeq
($a$ runs over ${\bf p}^{d-1}$) is a $(d-1)$-dimensional divergence-free vector. Here everything is evaluated at $\omega=0$. 

Class AIII, 
without $T$ or $C$ symmetry, is potentially nontrivial in any odd dimension. 
Classes BDI, CII, CI, and DIII are characterized by the presence of 
both $T$ and $C$ symmetries, with $\epsilon_T \epsilon_C=\eta$ equal to $+1$ 
in the first two classes and $-1$ in the other two. The 
invariant \rfs{eq:topinvchialt} vanishes if $\eta=1$ and $d=3+4n$, or 
if $\eta=-1$ and $d=1+4n$, because $\Sigma$ and $G$ anticommute at 
$\omega=0$ as a consequence of $S$-symmetry and $U_T \Sigma U_T^\dagger=\eta \Sigma^T$. 
Arguments similar to the ones employed above in nonchiral classes can now be 
used to see the emergence of the chiral ${\mathbb Z}_2$ invariant in 
appropriate dimensions. 

For example if $d=1$, the vector ${\bf r}$ reduces to a scalar
${r} = - \Tr \, \Sigma$. As is well known, zero-energy states are eigenstates 
of $\Sigma$, with the eigenvalue $+1$ 
(right zero modes) or $-1$ (left). 
$\Tr \, \Sigma$, the difference of the number of right and left 
eigenstates,
is then equal to the difference of the bulk invariants on both sides of this zero-dimensional boundary. 
A reader may worry that the formalism reported earlier in this article becomes somewhat degenerate at $d=1$, therefore we present a slightly more detailed derivation 
of this result in Appendix~\ref{sec:chiralone}. 

This concludes the derivation of \rfs{eq:theorem} for all classes of topological insulators in all spacial dimensions. 

\section{Conclusions}
In this article we presented a derivation of the relationship \rfs{eq:theorem} between the bulk and the boundary of topological insulators. 
Even when there are no interactions and no disorder, this relation is quite useful and allows to test whether a particular proposed boundary theory 
can indeed be at the edge of a topological insulator.  When interactions are present, the relationship \rfs{eq:theorem} remains true and still allows to relate the 
the bulk to the edge, although the edge states may get replaced by 
zeroes.~\cite{Gurarie2011}
In addition to the extension to interacting systems already described,
the method described here can also be extended to disordered topological insulators, by periodically repeating the finite
size system. The results obtained in this way will be reported elsewhere.


\acknowledgements
The authors thank A.~W.~W.~Ludwig and A.~Altland for many 
stimulating discussions in the course of this work. This work was 
supported by the NSF grants No.~DMR-0449521
and No.~PHY-0904017. VG would like to thank the Institute for Theoretical 
Physics at the University of Cologne for hospitality and the 
A.~v.~Humboldt Foundation for support during a sabbatical stay. We are also grateful to the Aspen Center for Physics where this work was completed. 

\appendix


\section{Derivation of Eq.~\rf{eq:rnconnection}. }
\label{sec:derivation}

Given a matrix $\tilde A(s,s')$ which is also a function of two variables $s$ and $s'$, one can
introduce its Wigner (or Weyl) transform
\be A \left(p_d, \bar s \right) = \int dr \, e^{i p_d r} \tilde A\left( \bar s + \frac r 2, \bar s - \frac r 2 \right).
\ee
The technical tool we need in the following calculation is the Moyal product expansion,
which allows us to write the Wigner transform of 
$\tilde C = \tilde A \circ \tilde B$ (this stands for $\tilde C(s,s'') = \int ds' \, A(s,s') B(s',s'')$) as
\be \label{eq:starexp}
C =  A  B + \frac{1}{2i} \epsilon_{\mu \nu} \, \d_\mu  A \, \d_\nu  B 
+ \dots,
\ee
where $\mu, \nu$ stand for $\bar s$, $p_d$.  In particular, this means that
\beq \label{eq:Moyal}  
{K} ={G}^{-1} +
\frac{1}{2i} \epsilon_{\mu \nu} \chi_\mu \chi_\nu {G}^{-1} + \dots, 
\quad \chi_\mu \equiv {G}^{-1} \d_\mu{G}.
\eeq
This can be checked by demanding that $\tilde K \circ \tilde G = 1$, applying
Wigner transform and using Eq.~\rf{eq:starexp}. 

The Moyal product expansion is a formal expansion in powers of the gradient 
of $G$, and the omitted terms involve higher gradients of ${G}$.  As
mentioned in the text, this does not limit the applicability of our results
to slow domain walls; since we are deriving a relation between integer-valued
quantities, any corrections from higher powers of the gradient must vanish
(unless they also evaluate to integers, in which case they would indicate
extra topological structure).

We would like to relate the vector ${\bf r}$ to the vector ${\bf n}$. The former involves 
Green's functions in real space, while the latter involves 
Wigner-transformed Green's functions. 
Therefore, we need to apply the gradient expansion to 
\beq
r_{a_0} = C_{d-2} \epsilon_{a_0 \dots a_{d-1}} 
\Tr\, X_{a_1} \circ \dots \circ X_{a_{d-1}}, \;\; 
X_{a} \equiv \tilde{G}^{-1} \circ \d_\mu \tilde{G} .
\eeq
First, consider the trace of a product:
\begin{align}
\label{eq:traceinvariance}
\Tr\, \tilde{A}\circ\tilde{B} 
&= \int\! dsds'\, \tr\,\tilde{A}(s,s')\tilde{B}(s',s) \notag\\
&= \int\! \frac{ds ds' dp_d dp'_d}{(2\pi)^2}\,  
e^{-ip_d(s-s') - ip'_d(s'-s)} \notag\\
&\hspace*{23mm} \tr\, A\left(p_d, \frac{s+s'}{2} \right)
B\left( p'_d,\frac{s+s'}{2}\right) \notag\\
&= \int\! \frac{d\bar{s}dp_d}{2\pi}\, \tr\, A(p_d,\bar{s}) B(p_d,\bar{s}) .
\end{align}
This states that the trace is invariant upon changing to the 
Wigner-Weyl basis.  The expansion we need is therefore
\beq
\Tr\, \tilde{K}\circ \d_{a_1} \tilde{G} \circ \dots \circ 
\tilde{K} \circ \d_{a_{d-1}} \tilde{G}
= \int\! \frac{d\bar{s}dp_d}{2\pi}\, \tr\, K M,
\eeq
where $M$ is the Wigner transform of 
$\d_{a_1} \tilde{G} \circ \tilde{K} \dots \circ \d_{a_{d-2}} \tilde{G} \circ \tilde{K} \circ \d_{a_{d-1}} \tilde{G}$:
\begin{widetext}
\begin{align} \label{eq:4}
M &= \left[ \left( \d_{a_1} G 
+ \frac{\epsilon_{\mu_1\nu_1}}{2i} \d_{\mu_1,a_1} G \d_{\nu_1}  \right) 
\left( K + \frac{\epsilon_{\mu'_1\nu'_1}}{2i} \d_{\mu'_1} K \d_{\nu'_1}  
\right) \right] \notag\\ &\quad\dots \left[ \left( 
\d_{a_{d-2}} G + \frac{\epsilon_{\mu_{d-2}\nu_{d-2}}}{2i} 
\d_{\mu_{d-2},a_{d-2}} G \d_{\nu_{d-2}}  \right) 
\left( K + \frac{\epsilon_{\mu'_{d-2}\nu'_{d-2}}}{2i} \d_{\mu'_{d-2}} K 
\d_{\nu'_{d-2}}  \right) \right] \d_{a_{d-1}} G + \dots \notag\\
&= \left[ \d_{a_1} G K 
+ \frac{\epsilon_{\mu_1\nu_1}}{2i} \d_{\mu_1,a_1} G \d_{\nu_1} (K\,\cdot) 
+ \frac{\epsilon_{\mu_1\nu_1}}{2i} \d_{\mu_1} K \d_{\nu_1} \right] \notag\\
&\quad \dots \left[ \d_{a_{d-2}} G K 
+ \frac{\epsilon_{\mu_{d-2}\nu_{d-2}}}{2i} \d_{\mu_{d-2},a_{d-2}} G \d_{\nu_{d-2}} (K\,\cdot) 
+ \frac{\epsilon_{\mu_{d-2}\nu_{d-2}}}{2i} \d_{a_{d-2}} G \d_{\mu_{d-2}} K \d_{\nu_{d-2}} \right] \d_{a_{d-1}} G + \dots .
\end{align}
\end{widetext}
Eq.~(\ref{eq:4}) comes from applying the Moyal product expansion, 
Eq.~(\ref{eq:starexp}), to every $\circ$-product. 

The notation is quite unwieldy, so we make the
following notational simplifications.  First, we suppress all
$\epsilon_{\mu\nu}$; since we will only be keeping terms of first order
in the gradient, there should be no confusion about how indices are
contracted.  Second, the factor $\epsilon_{a_0 \dots a_{d-1}}$ is
suppressed as well, and every other symbol that carries an index $a_i$ 
is written in bold face, with the index suppressed: 
$\chi_{a_i} \rightarrow \chib$ and $\d_{a_i} \rightarrow \dd$.  It must
be remembered that these symbols form a completely antisymmetric tensor.
In more formal language, these are one-forms.  We can also rewrite 
products (and $\circ$ products)
\beq
A_{a_1} \dots A_{a_j} \rightarrow \bf{A}^j.
\eeq
Finally, we will write $d-1 = n$ (which is odd by assumption).

Then the quantity to be computed, up to the overall constant, is
\begin{widetext}
\begin{align}
&\Tr\, \X^n = \int\! \frac{d\bar{s}dp_d}{2\pi}\, \tr\, K
\left[ \dd G K + \frac{1}{2i} \d_{\mu} \dd G \d_{\nu} (K\,\cdot) 
+ \frac{1}{2i} \dd G \d_{\mu} K \d_{\nu} \right]^{n-1} \!\dd G + \dots .
\end{align}
The bracket expands to
\begin{align}
(\dd G K)^{n-1} + \frac{1}{2i} \sum_{j=0}^{n-2} 
(\dd G G^{-1})^{n-2-j} \left[ \d_{\mu} \dd G \d_{\nu} (G^{-1}\,\cdot) 
+ \dd G \d_{\mu} G^{-1} \d_{\nu} \right] (\dd G G^{-1})^{j} +\dots,
\end{align}
where $K \rightarrow G^{-1}$ in the sum because we keep only
first order in the gradient.  Then
\begin{align}
\Tr\, \X^n &= \int\! \frac{d\bar{s}dp_d}{2\pi}\, \tr\, \Bigg\{ (K\dd G)^n
\notag\\ &\quad + \frac{1}{2i} \sum_{j=0}^{n-2} 
(G^{-1}\dd G )^{n-2-j} G^{-1}\left[ \d_{\mu} \dd G \d_{\nu} (G^{-1}\,\cdot) 
+ \dd G \d_{\mu} G^{-1} \d_{\nu} \right] \dd G ( G^{-1}\dd G)^{j} \Bigg\} 
+ \dots \notag\\
&= \int\! \frac{d\bar{s}dp_d}{2\pi}\, \Bigg\{ \tr\, \chib^n
+ \frac{1}{2i} \sum_{j=1}^{n} \tr\, \chib^{n-j} \chi_\mu \chi_\nu \chib^j
\notag\\ &\quad + \frac{1}{2i} \sum_{j=0}^{n-2} 
\tr\, \chib^{n-2-j} \left[ G^{-1}\d_{\mu} \dd G \d_{\nu} (\chib^{j+1}) 
- \chib \chi_\mu G^{-1} \d_{\nu}(G \chib^{j+1}) \right] \Bigg\} +\dots,
\end{align}
\end{widetext}
where we have expanded $K$ in the first term according to Eq.~\eqref{eq:Moyal} and replaced 
$\chib = G^{-1} \dd G$ and $\d_\mu G^{-1} = -\chi_\mu G^{-1}$.

Note that the the zeroth-order term in the expansion,
\beq
\int\! \frac{d\bar{s}dp_d}{2\pi}\, \tr\, \chib^n,
\eeq
produces what is often termed a ``weak'' topological invariant,
\aem{refs}
that is,
a topological invariant of lower dimension than the bulk system; we will
ignore this term from now on.

Because $n$ is odd, no sign is picked up under
cyclic relabeling of the indices hidden in the bold-face notation, so
we can use cyclicity of the trace.  Therefore,
we can combine terms to obtain the first order quantity
\begin{align}
\bf{T} &\equiv \sum_{j=1}^{n} \tr\, \chib^{n-j} \chi_\mu \chi_\nu \chib^j
+ \sum_{j=0}^{n-2} 
\tr\, \chib^{n-2-j} \notag\\
&\hspace*{10mm}\times\left[ G^{-1}\d_{\mu} \dd G \d_{\nu} (\chib^{j+1}) 
- \chib \chi_\mu G^{-1} \d_{\nu}(G \chib^{j+1}) \right] \notag\\
&= \tr\, \chib^{n} \chi_\mu \chi_\nu \notag\\
&\qquad{}+ \sum_{j=0}^{n-2} \tr\, \chib^{n-2-j} 
\left[ G^{-1}\d_\mu \dd G - \chib \chi_\mu \right] \d_\nu (\chib^{j+1}) .
\notag
\intertext{The combination in square brackets is simply $\dd \chi_\mu$ so,
shifting the summation index,}
&= \tr\, \chib^{n} \chi_\mu \chi_\nu
+ \sum_{j=1}^{n-1} \tr\, \chib^{n-1-j} (\dd \chi_\mu) \d_\nu (\chib^{j}).
\end{align}

The factor $\dd \chi_\mu$ contains double derivatives of $G$,
which do not appear in our desired result.  Therefore, we will separate
out total derivatives in $\dd$, which produce no flux on the (closed)
Gaussian surfaces $S_F$ on which the topological invariant is built [see Eq.~\eqref{eq:rnconnection}].
To do this, note that $\dd^2 = 0$ and that, when using the product rule
to expand these derivatives, a sign appears every time $\dd$ moves past 
a $\chib$.  This gives
\begin{align}
\dd(\chib^{j}) 
&= \sum_{j'=0}^{j-1} (-1)^{j'} \chib^{j'} (\dd\chib) \chib^{j-1-j'} \notag\\
&= \sum_{j'=0}^{j-1} (-1)^{j'+1} \chib^{j+1}
= \begin{cases} 
0 & j\,\text{even} \\ -\chib^{j+1} & j\,\text{odd} \end{cases} ,
\end{align}
which implies
\begin{widetext}
\begin{align}
\tr\, \chib^{n-j-1} (\dd \chi_\mu) \d_\nu (\chib^j) = 
\begin{cases}
\dd \,\tr\, \chib^{n-j-1} \chi_\mu \d_\nu (\chib^j) & j\,\text{even} \\
-\dd \,\tr\, \chib^{n-j-1} \chi_\mu \d_\nu (\chib^j) 
- \tr\, \chib^{n-j} \chi_\mu \d_\nu (\chib^j)
+ \tr\, \chib^{n-j-1} \chi_\mu \d_\nu (\chib^{j+1})
&j\,\text{odd}
\end{cases}
\end{align}
\end{widetext}
since $n$ is odd.  Then
\begin{align}
\bf{T} &= \tr\, \chib^{n} \chi_\mu \chi_\nu
+ \dd \sum_{j=1}^{n-1} (-1)^j \tr\, \chib^{n-j-1} \chi_\mu \d_\nu (\chib^j)
\notag\\
&\quad{}+ \sum_{\begin{subarray}{c}j=1\\j\,\mathrm{odd}\end{subarray}}^{n-1}
\left[ \tr\, \chib^{n-j-1} \chi_\mu \d_\nu (\chib^{j+1})
- \tr\, \chib^{n-j} \chi_\mu \d_\nu (\chib^j) \right] \notag\\
&= \dd \sum_{j=1}^{n-1} (-1)^j \tr\, \chib^{n-j-1} \chi_\mu \d_\nu (\chib^j)
+ \tr\, \chib^{n} \chi_\mu \chi_\nu \notag\\
&\quad{}+ \sum_{j=1}^{n-1} (-1)^j \tr\, \chib^{n-j} \chi_\mu \d_\nu (\chib^j),
\end{align}
where we have shifted the summation index in the last term.  The total 
derivative is of no further interest, so we set 
$\bf{D}_1 = 
\dd \sum_{j=1}^{n-1} (-1)^j \tr\, \chib^{n-j-1} \chi_\mu \d_\nu (\chib^j)$.  
In the last term, expanding the derivative $\d_\nu$ and using cyclicity 
gives
\begin{align}
\bf{T} &= \bf{D}_1 + \tr\, \chib^{n} \chi_\mu \chi_\nu \notag\\
&\quad{}+ \sum_{j=1}^{n-1} \sum_{m=1}^j (-1)^j \tr\, \chib^{n-m} \chi_\mu 
\chib^{m-1} \d_\nu \chib \notag\\
&= \bf{D}_1 + \tr\, \chib^{n} \chi_\mu \chi_\nu \notag\\
&\quad{}+ \sum_{m=1}^{n-1} \left[ \sum_{j=m}^{n-1} (-1)^j \right]
\tr\, \chib^{n-m} \chi_\mu \chib^{m-1} \d_\nu \chib \notag\\
&= \bf{D}_1 + \tr\, \chib^{n} \chi_\mu \chi_\nu
+ \sum_{\begin{subarray}{c}m=2\\m\,\mathrm{even}\end{subarray}}^{n-1}
\tr\, \chib^{n-m} \chi_\mu \chib^{m-1} \d_\nu \chib . \notag
\intertext{With 
$\d_\nu \chib = \dd \chi_\nu + \chib \chi_\nu - \chi_\nu \chib$,}
&= \bf{D}_1 + \tr\, \chib^{n} \chi_\mu \chi_\nu
+ \sum_{\begin{subarray}{c}m=2\\m\,\mathrm{even}\end{subarray}}^{n-1}
\tr\, \chib^{n-m} \chi_\mu \chib^{m-1} \dd \chi_\nu \notag\\
&\quad{}+ \!\!\sum_{\begin{subarray}{c}m=2\\m\,\mathrm{even}\end{subarray}}^{n-1} \!\!
\left[ \tr \chib^{n-m} \chi_\mu \chib^{m} \chi_\nu - \tr \chib^{n-m+1} \chi_\mu \chib^{m-1} \chi_\nu \right].
\end{align}
Now, because $n$ is odd and $m$ is even,
\begin{align}
&\sum_{\begin{subarray}{c}m=2\\m\,\mathrm{even}\end{subarray}}^{n-1}
\dd \,\tr\, \chib^{n-m} \chi_\mu \chib^{m-1} \chi_\nu \notag\\
&\;= 
\sum_{\begin{subarray}{c}m=2\\m\,\mathrm{even}\end{subarray}}^{n-1}
\left[
- \,\tr\, \chib^{n-m+1} \chi_\mu \chib^{m-1} \chi_\nu \right.\notag\\
& \qquad\qquad \left. {}
- \tr\, \chib^{n-m} (\dd \chi_\mu) \chib^{m-1} \chi_\nu \right.\notag\\
& \qquad\qquad \left. {}+ \tr\, \chib^{n-m} \chi_\mu \chib^{m} \chi_\nu 
+ \tr\, \chib^{n-m} \chi_\mu \chib^{m-1} \dd \chi_\nu \right] \notag\\
&= 
\sum_{\begin{subarray}{c}m=2\\m\,\mathrm{even}\end{subarray}}^{n-1}
\left[
2 \,\tr\, \chib^{n-m} \chi_\mu \chib^{m-1} \dd \chi_\nu 
+ \tr\, \chib^{n-m} \chi_\mu \chib^{m} \chi_\nu \right.\notag\\
& \qquad\qquad \left. {}
- \tr\, \chib^{n-m+1} \chi_\mu \chib^{m-1} \chi_\nu \right], 
\end{align}
where the second equality follows from antisymmetry in $\mu$ and $\nu$
after a relabeling of the summation index in the second term of the
first equality.  This means
\begin{align}
\bf{T} &= \bf{D}_1 + \tr\, \chib^{n} \chi_\mu \chi_\nu \notag\\
&\quad{}+ \frac{1}{2} 
\sum_{\begin{subarray}{c}m=2\\m\,\mathrm{even}\end{subarray}}^{n-1}
\{ \dd \,\tr\, \chib^{n-m} \chi_\mu \chib^{m-1} \chi_\nu \notag\\
&\hspace*{20mm}+ \tr\, \chib^{n-m} \chi_\mu \chib^{m} \chi_\nu \notag\\
&\hspace*{20mm}- \tr\, \chib^{n-m+1} \chi_\mu \chib^{m-1} \chi_\nu \}.
\end{align}
We can collect the total derivatives as $\bf{D}_2$, shift the summation
index in the last term, and rewrite $2 \,\tr\, \chib^n \chi_\mu \chi_\nu = 
\tr\, \chib^n \chi_\mu \chi_\nu - \tr\, \chi_\mu \chib^n \chi_\nu $ to 
obtain
\begin{align}
\bf{T} &= \bf{D}_2 + \frac{1}{2} 
\sum_{m=0}^{n} (-1)^{m} \,\tr\, \chib^{n-m} \chi_\mu \chib^m \chi_\nu.
\end{align}
Including $\bf{D}_2$ with the suppressed terms and restoring 
$\epsilon_{\mu\nu}$, this means that
\beq
\Tr\, \X^n = \int\! \frac{d\bar sdp_d}{8\pi i} \sum_{m=0}^{n} (-1)^{m}
\epsilon_{\mu\nu} \tr\, \chib^{n-m} \chi_\mu \chib^m \chi_\nu + \dots.
\eeq

This expression is sufficiently simple that we can return to the 
original, more explict notation:
\begin{align}
&\epsilon_{a_0 a_1 \dots a_n} \Tr\, X_{a_1} \circ \dots \circ X_{a_n} \notag\\
&\;= \int\! \frac{d\bar sdp_d}{8\pi i} \sum_{m=0}^{n} (-1)^{m} 
\epsilon_{\mu\nu} \epsilon_{a_0 a_1 \dots a_n} \notag\\
&\hspace*{6mm}\times\tr\, \chi_{a_1} \dots \chi_{a_{n-m}} \chi_\mu 
\chi_{a_{n-m+1}} \dots \chi_{a_{n}} \chi_\nu +\dots.
\end{align}
The right-hand side has $2n!(n+1) = 2 (n+1)!$ terms and is totally 
antisymmetric, while using the Levi-Civita symbol gives $(n+2)!$ terms.
Therefore,
\begin{align}
&\epsilon_{a_0 a_1 \dots a_n} \Tr\, X_{a_1} \circ \dots \circ X_{a_n} \notag\\
&\;= \frac{1}{4\pi i (n+2)} \int\! d\bar sdp_d \, \epsilon_{a_0 a_1 \dots a_{n+2}} 
\tr\, \chi_{a_1} \dots \chi_{a_{n+2}} + \dots,
\end{align}
assuming that we choose to order the new indices appropriately.

Finally, it is important that the constant be correct:
\begin{align}
C_{d} &= -(2\pi i)^{-\frac{d}{2}-1}\frac{(d/2)!}{(d+1)!} \notag\\
&= (2\pi i)^{-1} \frac{d/2}{d(d+1)} \left[ -(2\pi i)^{-\frac{d}{2}}\frac{(d/2 - 1)!}{(d-1)!} \right] \notag\\
&= \frac{1}{4\pi i(d+1)} C_{d-2},
\end{align}
so (substituting back $d=n+1$)
\begin{align}
&C_{d-2}\epsilon_{a_0 a_1 \dots a_{d-1}} 
\Tr\, X_{a_1} \circ \dots \circ X_{a_{d-1}} \notag\\
&\;= \frac{1}{4\pi i (d+1)} C_{d-2} \!\!
\int\!\! d\bar sdp_d \, \epsilon_{a_0 a_1 \dots a_{d+1}} 
\tr \chi_{a_1} \dots \chi_{a_{d+1}} \notag\\ &\hspace*{75mm}{}+ \dots \notag\\
&\;= C_{d} \int\! d\bar sdp_d \, \epsilon_{a_0 a_1 \dots a_{d+1}} 
\tr\, \chi_{a_1} \dots \chi_{a_{d+1}} +\dots,
\end{align}
or
\beq \label{eq:mainrelation}
r_{a} = \int\! d\bar sdp_d \, n_{a} + \dots.
\eeq
This is valid for values of $a$ in $(\omega, {\bf p}^{d-1})$, but does not 
make sense for $a=\bar s$ or $a=p_d$ since ${\bf r}$ does not have such 
components. 

We would like to calculate 
\be \int d{\bf S}^{d+1}_f \cdot {\bf n}.
\ee
We deform the sphere $S_f$ to the space $S_F \times M$ where $S_F$ is the sphere in the space
$(\omega, {\bf p}^{d-1})$ and
$M$ is the entire space (${\mathbb R}^2$) spanned by $p_d$ and $\bar s$. This can
be termed a ``hypercylinder," since a cylinder is the product of a circle (analogous to our sphere $S_F$) and a straight line (analogous to the flat
space $M = {\mathbb R}^2$).

The flux of ${\bf n}$ through such a surface can be computed with only the components of ${\bf n}$ 
for which the relation Eq.~\rf{eq:mainrelation} holds. Part of that flux involves integration over
$ds dp_d$, which then naturally leads to
\be  
\int\! d{\bf S}^{d+1}_f \cdot {\bf n} 
= \int\! d{\bf S}^{d-1}_F \cdot \int\! d\bar s dp_d \,{\bf n} 
= \int d{\bf S}^{d-1}_f \cdot {\bf r},
\ee
where \rfs{eq:mainrelation} was used. This is the result asserted in the text as Eq.~\eqref{eq:rnconnection}.


\section{Chiral systems in one dimensional space}
\label{sec:chiralone}
We start by writing $\Tr \, \Sigma$ as
\be \Tr \, \Sigma =  \Tr \, \Sigma \, \tilde K \circ \tilde G.
\ee
Notice that this is true at arbitrary $\omega$. 
We then take advantage of Eqs.~\rf{eq:Moyal} and \rf{eq:traceinvariance} to rewrite this as
\be \frac{1}{4\pi i}  \tr \Sigma \int d\bar s dp  \left( G^{-1} \d_{\bar s} G G^{-1} \d_p G - G^{-1} \d_p G G^{-1} \d_{\bar s} G \right).
\ee
The expression to be integrated is a total derivative and results, upon integrating, in
\be \Tr \, \Sigma = N_1(s=-L) - N_1(s=L),
\ee
where $N_1$, defined in \rfs{eq:topinvchialt}, is evaluated at  $\bar s=-L$ and $\bar s=L$.
This is indeed what is claimed in section \ref{sec:chiral}. What remains is to remark that while these expressions are written at arbitrary $\omega$, the limit $\omega
\rightarrow 0$ is convenient as only in this limit $\Tr \, \Sigma$ counts zero energy states  (and zeroes of Green's functions if there are interactions) localized at the boundary. 
This derivation can be used instead of the much more involved procedure reported in Ref.~\onlinecite{Gurarie2011}.

\bibliography{shortbib}

\end{document}